\documentclass[conference]{IEEEtran}
\IEEEoverridecommandlockouts
\usepackage{cite}
\usepackage{amsmath,amssymb,amsfonts}
\usepackage{algorithmic}
\usepackage{graphicx}
\usepackage{balance}
\usepackage{textcomp}
\usepackage[caption=false,font=footnotesize]{subfig}
\usepackage{xcolor}

\DeclareMathOperator{\EX}{\mathbb{E}}
\DeclareMathOperator{\tr}{tr} 
\DeclareMathAlphabet\mathbfcal{OMS}{cmsy}{b}{n} 
\DeclareMathOperator*{\argmin}{arg\,min}
\newcommand{\RNum}[1]{\lowercase\expandafter{\romannumeral #1\relax}}

\usepackage[linesnumbered,ruled,vlined]{algorithm2e} 
\makeatletter
\newcommand{\nosemic}{\renewcommand{\@endalgocfline}{\relax}}
\newcommand{\dosemic}{\renewcommand{\@endalgocfline}{\algocf@endline}}
\let\oldnl\nl
\newcommand{\nonl}{\renewcommand{\nl}{\let\nl\oldnl}}
\makeatother
\newcommand\blfootnote[1]{%
	\begingroup
	\renewcommand\thefootnote{}\footnote{#1}%
	\addtocounter{footnote}{-1}%
	\endgroup
}

\newcommand*\squeezespaces[1]{
	\thickmuskip=\scalemuskip{\thickmuskip}{#1}%
	\medmuskip=\scalemuskip{\medmuskip}{#1}%
	\thinmuskip=\scalemuskip{\thinmuskip}{#1}%
	\nulldelimiterspace=#1\nulldelimiterspace
	\scriptspace=#1\scriptspace
}
\newcommand*\scalemuskip[2]{%
	\muexpr #1*\numexpr\dimexpr#2pt\relax\relax/65536\relax
} 

\def\BibTeX{{\rm B\kern-.05em{\sc i\kern-.025em b}\kern-.08em
    T\kern-.1667em\lower.7ex\hbox{E}\kern-.125emX}}
\begin{document}
	\bstctlcite{IEEEexample:BSTcontrol}
\title{Sub-Array Selection in Full-Duplex Massive MIMO for Enhanced Self-Interference Suppression \vspace{-1ex}}
\author{Mobeen Mahmood, Asil Koc, Duc Tuong Nguyen, Robert Morawski, Tho Le-Ngoc \\ Department of Electrical and Computer Engineering, McGill University, Montreal, QC, Canada \\
	Email: mobeen.mahmood@mail.mcgill.ca, asil.koc@mail.mcgill.ca, tuong.nguyen2@mail.mcgill.ca,\\ robert.morawski@mcgill.ca, tho.le-ngoc@mcgill.ca \vspace{-3ex}}

\maketitle

\begin{abstract}
	This study considers a novel full-duplex (FD) massive multiple-input multiple-output (mMIMO) system using hybrid beamforming (HBF) architecture, which allows for simultaneous uplink (UL) and downlink (DL) transmission over the same frequency band. Particularly, our objective is to mitigate the strong self-interference (SI) solely on the design of UL and DL RF beamforming stages jointly with sub-array selection (SAS) for transmit (Tx) and receive (Rx) sub-arrays at base station (BS). Based on the measured SI channel in an anechoic chamber, we propose a min-SI beamforming scheme with SAS, which applies perturbations to the beam directivity to enhance SI suppression in UL and DL beam directions. To solve this challenging non-convex optimization problem, we propose a swarm intelligence-based algorithmic solution to find the optimal perturbations as well as the Tx and Rx sub-arrays to minimize SI subject to the directivity degradation constraints for the UL and DL beams. The results show that the proposed min-SI BF scheme can achieve SI suppression as high as 78 dB in FD mMIMO systems.
	\vspace{-1em}
\end{abstract}

\blfootnote{This work was supported in part by Huawei Technologies Canada and in part by the Natural Sciences and Engineering Research Council of Canada.}

\IEEEpeerreviewmaketitle
\vspace{-1em}
\section{Introduction}
\vspace{-1ex}
\IEEEPARstart{T}{he} ever-increasing demand for data traffic has presented a considerable challenge for future wireless communications systems, which must efficiently utilize the available frequency spectrum. In this regard, the full-duplex (FD) communications technology has demonstrated potential for significant improvement in spectral efficiency as compared to traditional frequency and time-division duplexing systems. The simultaneous transmission of uplink (UL) and downlink (DL) signals in the same frequency and time resources in FD communications has the potential to theoretically double the capacity by utilizing resources effectively \cite{FD_2}. Massive multiple-input multiple-output (mMIMO), which is a pivotal enabler of fifth-generation (5G) networks, utilizes large array structures at the base station (BS) to serve multiple users via spatial multiplexing. The three dimensional (3D) beamforming of mMIMO can further enhance the performance by exploiting the additional spatial degrees of freedom (DoF) offered by multiple transmitter (Tx) and receiver (Rx) antennas. Thus, FD and mMIMO together can fulfill the throughput and latency demands of 5G and beyond 5G (B5G) wireless communications systems with limited spectrum resources \cite{FD_mMIMO}. \par 
The simultaneous transmission and reception of FD communications over the same frequency band may sound like a promising solution, but it comes with a serious challenge: strong self-interference (SI). Contrary to half-duplex (HD) communications, SI, which is produced as a result of the strong transmit signal's coupling with the Rx chains, has a significant adverse effect on the performance of FD systems because it can impair the Rx antennas' ability to receive the UL signal. Many research efforts have focused on SI suppression in FD systems to fully utilize this technology \cite{FD_SIC_1,FD_SIC_2,FD_SIC_3}. In particular, different SI suppression techniques can be broadly classified as follows: 1) antenna isolation; 2) analog cancellation; and 3) digital cancellation \cite{SIC_antenna_3,SIC_analog_1,SIC_digital_1}. In FD communications systems, antenna isolation, analog/digital SI cancellation (SIC), and their combinations have been used to effectively suppress the strong SI signal below the Rx noise level \cite{SIC_joint_1}.\par

In 5G and B5G systems, there is a growing trend toward utilizing an increased number of antennas at BS. For instance, the third generation partnership project (3GPP) has been contemplating the deployment of 64-256 antenna configurations \cite{3GPP_mMIMO}.  However, this poses a significant hurdle for analog SIC in FD mMIMO systems, as the associated analog complexity becomes prohibitively large as an increased number of antennas results in more SI components. To mitigate this challenge, SoftNull relies exclusively on transmit beamforming to mitigate SI, thereby completely obviating the need for analog cancelers \cite{SIC_mMIMO_digital}. In the realm of mMIMO HD systems, fully-digital beamforming (FDBF) and hybrid beamforming (HBF) are two common approaches for mitigating interference. Recent studies, for instance, SoftNull, have exploited the availability of multiple antennas in FD mMIMO systems in order to provide SI suppression via FDBF, commonly referred to as spatial suppression. However, FDBF becomes infeasible for mMIMO systems with very large array structures due to the following reasons: 1) prohibitively high cost; 2) complexity; and 3) energy consumption. Conversely, HBF, which involves the design of both the radio frequency (RF) and baseband (BB) stages, can approach the performance of FDBF by reducing the number of energy-intensive RF chains, thereby minimizing power consumption. \par 
In the related studies, various HBF techniques, which include HD transmission in DL and UL, are investigated in \cite{koc2020Access,mahmood2021energy,mobeen_3D} and for FD transmission in \cite{koc2021,FD_Precoding_4,Koc2022_FD_MU,FD_Precoding_5,FD_Precoding_7}. In particular, the authors in \cite{koc2020Access,mahmood2021energy,mobeen_3D} introduce different HBF techniques, where the RF stage is constructed utilizing users' angular information only. The angle-of-departure (AoD) and angle-of-arrival (AoA) information is used in \cite{koc2021} to propose a hybrid precoding/combining (HPC) technique for a millimeter-wave (mmWave) FD mMIMO system to suppress SI and decrease the number of RF chains. The authors in \cite{FD_Precoding_4} introduced the HPC for an FD amplify-and-forward (AF) relay using correlated estimation errors to mitigate SI. For the multi-user (MU) FD mMIMO system in \cite{Koc2022_FD_MU}, the non-orthogonal beams are generated to serve multiple users to maximize sum-rate capacity while suppressing the strong SI. Similarly, the authors in \cite{FD_Precoding_5} show that SI can be reduced by around 30 dB through the joint design of the transmit and receive RF beamformer weights, as well as the precoder and combiner matrices. A two-timescale HBF scheme for FD mmWave multiple-relay transmission is investigated in \cite{FD_Precoding_7}, where the analog and digital beams are updated based on channel samples and real-time low-dimensional effective channel state information (CSI) matrices, respectively. Most hybrid mMIMO systems consider either fully-connected (FC) or sub-array-connected (SAC) HBF architectures. Compared to FC, SAC requires a lower number of phase shifters (PSs). Thus, its use can reduce power consumption at the expense of some performance degradation; however, it can provide a better spectral-energy efficiency tradeoff \cite{Sub-array1}. Compared to HD transmissions, the use of SAC in FD mMIMO systems is limited. The use of SAC architecture both for Tx and Rx in FD transmissions can provide an additional DoF to suppress strong SI.\par 
To address this gap in the literature, this paper introduces a novel sub-array selection scheme (SAS) to suppress strong SI in FD mMIMO systems using a measured SI channel. To the best of our knowledge, this is the first work that considers SI suppression solely based on the design of the transmit and receive RF beamforming stages and SAS. Our objective here is to show that the SI level can be reduced to the noise floor by merely employing the spatial DoF that large array architectures afford, without the need for expensive and complicated analogue cancellation circuits. We propose a novel min-SI hybrid beamforming scheme, which applies perturbations to the beam directions jointly with SAS for enhanced SI suppression. To reduce the high computational complexity during the search for optimal perturbations, we propose a swarm intelligence-based algorithmic solution to find the optimal perturbations, and the Tx/Rx sub-arrays to minimize SI while satisfying the directivity degradation constraints for the UL and DL beams. The results show that the proposed min-SI BF scheme together with SAS can achieve SI suppression as high as 78 dB in real-time implementations.
\begin{figure}[!t]
	\centering
	\captionsetup{justification=centering}
	\includegraphics[height= 5cm, width=0.9\columnwidth]{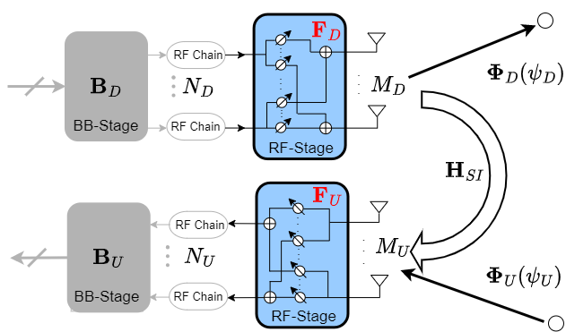} 
	\caption{System model of FD mMIMO HBF communications system.}
	\label{fig:fig1}
	\vspace{-3em}
\end{figure} 
\begin{figure}[!t]
	\centering
	\captionsetup{justification=centering}
	\includegraphics[height= 4.5cm, width=1\columnwidth]{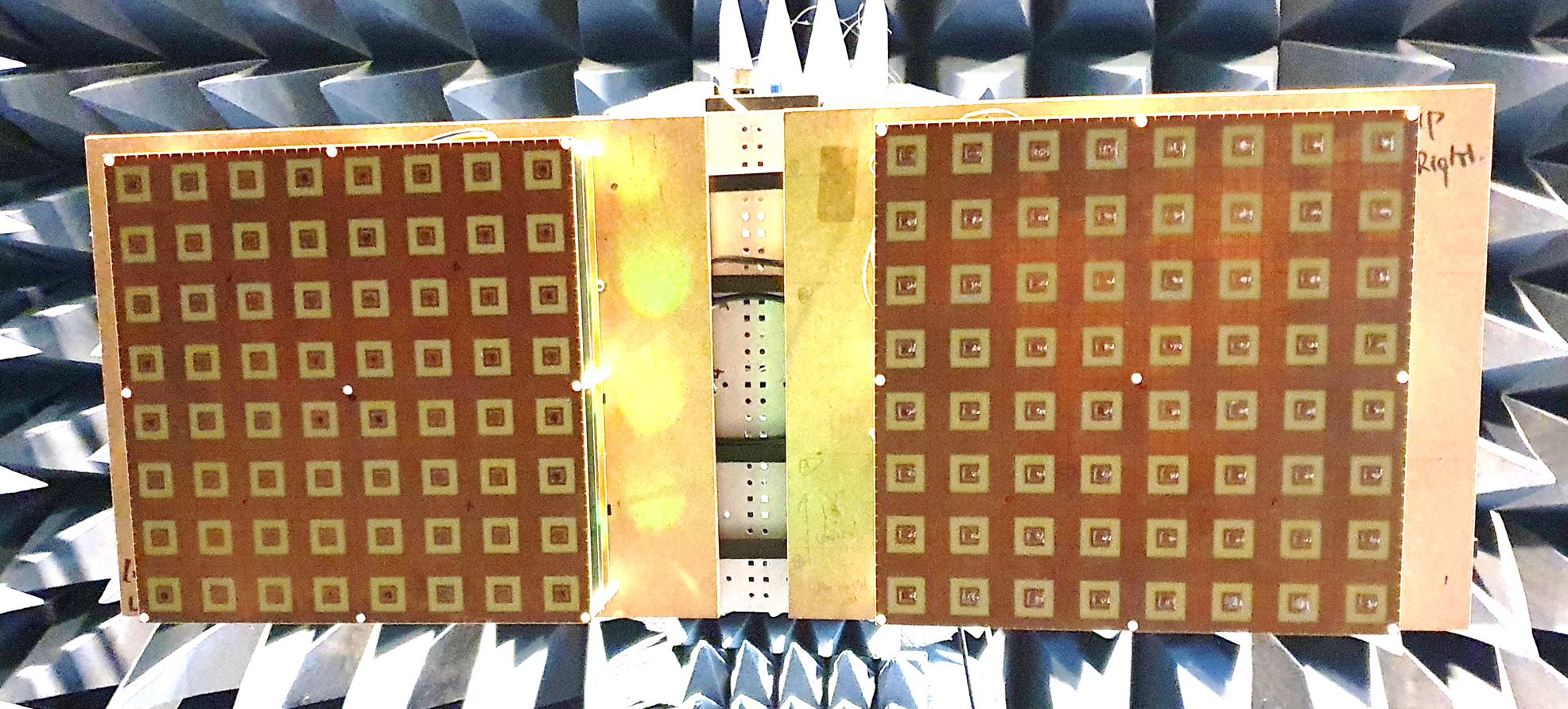} 
	\caption{Tx and Rx antenna setup in anechoic chamber.}
	\label{fig:fig2}
	\vspace{-2em}
\end{figure} 
\section{System Model \& Measured SI Channel}
\subsection{System Model}
We consider a single-cell FD mMIMO system for joint DL and UL transmission as shown in Fig. \ref{fig:fig1}. Here, the BS operates in FD mode to simultaneously serve $K_D$ DL and $K_U$ UL single-antenna UEs over the same frequency band, while the UEs operate in HD mode due to the hardware/software constraints on UEs (e.g., low power consumption, limited signal processing and active/passive SI suppression capability). As shown in Fig. \ref{fig:fig2}, the BS is equipped with transmit/receive uniform rectangular arrays (URAs), which are separated by an antenna isolation block for passive (i.e., propagation domain) SI suppression. Specifically, the transmit (receive) URA has $M_D=M_D^{(x)} \times M_D^{(y)} (M_U=M_U^{(x)} \times M_U^{(y)})$ antennas, where $M_D^{(x)} (M_U^{(x)})$ and $M_D^{(y)}(M_U^{(y)})$ denote the number of transmit (receive) antennas along $x$-axis and $y$-axis, respectively.  \par 
For the proposed FD mMIMO system, we consider the DL signal is processed through DL BB stage $\mathbf{B}_D \in \mathbb{C}^{N_D \times K_D}$ and DL RF beamformer $\mathbf{F}_D \in \mathbb{C}^{M_D \times N_D}$, where $N_D$ is the number of RF chains such that $K_D \leq N_D \ll M_D$. Similarly, the received UL signal at BS is processed through UL RF beamformer $\mathbf{F}_U \in \mathbb{C}^{N_U \times M_U}$ and UL BB combiner $\mathbf{B}_U \in \mathbb{C}^{K_U \times N_U}$ by utilizing $K_U \leq N_U \ll M_U$ RF chains. Here, the UL and DL RF beamforming stages (i.e., $\mathbf{F}_U$ and $\mathbf{F}_D$) are built using low-cost PSs. The DL channel matrix is denoted as $\mathbf{H}_D \in \mathbb{C}^{K_D \times M_D}$ with $\mathbf{h}_{D, k} \in \mathbb{C}^{M_D}$ as the $k^{th}$ DL UE channel vector. Similarly, $\mathbf{H}_U \in \mathbb{C}^{M_U \times K_{U}}$ is the UL channel matrix with $\mathbf{h}_{U, k} \in \mathbb{C}^{M_U}$ as the $k^{th}$ UL UE channel vector. Due to the FD transmission, the SI channel matrix $\mathbf{H}_{S I} \in \mathbb{C}^{M_U \times M_D}$ is present between Tx and Rx antennas at the BS. For the DL transmission, the transmitted signal vector at the BS is defined as $\mathbf{s}_D=\mathbf{F}_D \mathbf{B}_D \mathbf{d}_D \in \mathbb{C}^{M_D}$, where $\mathbf{d}_D=\left[d_{D, 1}, \cdots, d_{D, K_D}\right]^T \in \mathbb{C}^{K_D}$ is the DL data signal vector such that $\EX\{\mathbf{d}_D \mathbf{d}_D^H\}=\mathbf{I}_{K_D}$. The transmitted signal vector satisfies the maximum DL transmit power constraint, which is $\EX\{||\mathbf{s}_D||^2\}=\tr(\mathbf{F}_D \mathbf{B}_D \mathbf{B}_D^H \mathbf{F}_D^H) \leq P_D$, where $P_D$ is the total DL transmit power. Then, the received DL signal vector is given as follows: 
\begin{equation}
	\mathbf{r}_D=\underbrace{\mathbf{H}_D \mathbf{F}_D \mathbf{B}_D \mathbf{d}_D}_{\text{Desired Signal}}+ \underbrace{\mathbf{H}_{U}\mathbf{d}_U}_{\text{IUI by UL UE}}  + \underbrace{\mathbf{w}_D}_{\text{Noise}},
\end{equation}
where $\mathbf{H}_U \in \mathbb{C}^{K_D \times K_U}$ is the inter-user interference (IUI) between the DL/UL UE and $\mathbf{w}_D=\left[w_{D, 1}, \cdots, w_{D, K_D}\right]^T$ $\sim$ $\mathcal{C N}\left(0, \sigma_W^2 \mathbf{I}_{K_D}\right)$ is the complex circularly symmetric Gaussian noise vector. Here, we define $P_U$ as the transmit power of each UL UE. Similar to the DL data signal vector, the UL received signal at BS can be written as follows:
\begin{equation}
	\tilde{\mathbf{r}}_U\hspace{-0.5ex}=\hspace{-0.5ex}\underbrace{\hspace{-0.35ex}\mathbf{B}_U\hspace{-0.35ex} \mathbf{F}_U\hspace{-0.35ex} \mathbf{H}_U\hspace{-0.35ex} \mathbf{d}_U\hspace{-0.35ex}}_{\text{Desired Signal}} + \underbrace{\hspace{-0.35ex}\mathbf{B}_{U}\hspace{-0.35ex}\mathbf{F}_{U}\hspace{-0.35ex}\mathbf{H}_{SI}\hspace{-0.35ex}\mathbf{F}_{D}\mathbf{B}_{D}\hspace{-0.15ex}\mathbf{d}_{D}\hspace{-0.15ex}}_{\text{SI}}\hspace{-0.5ex} + \hspace{-3ex}\underbrace{\tilde{\mathbf{w}_U}}_{\text{Modified Noise}},
\end{equation}
where $\mathbf{d}_U=\left[d_{U, 1}, \cdots, d_{U, K_U}\right]^T \in \mathbb{C}^{K_U}$ is the UL data signal vector such that $\mathrm{E}\left\{\mathbf{d}_U \mathbf{d}_U^H\right\}=\mathbf{I}_{K_U}$ and $\tilde{\mathbf{w}_U}=\mathbf{B}_U \mathbf{F}_U\mathbf{w}_U$, where $\mathbf{w}_U = \left[w_{u, 1}, \cdots, w_{U, K_U}\right]^T \sim$ $\mathcal{C N}(0, \sigma_W^2 \mathbf{I}_{K_U})$ is the complex circularly symmetric Gaussian noise vector. The desirable DL (UL) beam direction has azimuth and elevation angles $\theta_D (\theta_U)$ and $\psi_D (\psi_U)$, respectively. For simplicity, we consider the following: 1) a single UL and DL UE (i.e., $K_D = K_U = 1$) \footnote{For simplicity in presentation, in the following discussion, we consider a simple scenario of single UL and a single DL UE. However, the proposed scheme can be applied to multiple UL and DL UEs, which is left as our future work.}; and 2) a uniform linear sub-array, where $\psi_D=\psi_U=90^{\circ}$. Then, the phase response vectors of the DL and UL directions can be written as follows:
\begin{eqnarray}
	\mathbf{\Phi\hspace{-0.25ex}}_D \hspace{-0.25ex}(\hspace{-0.25ex}\theta_D\hspace{-0.25ex}) \hspace{-0.75ex}& \hspace{-1ex}=\hspace{-1ex}&\hspace{-0.15ex} \mbox{$\squeezespaces{0.1}\hspace{-0.75ex} [\hspace{-0.25ex}1\hspace{
			-0.2ex},\hspace{-0.25ex}e^{\hspace{-0.25ex}- j2\pi d {{\cos (\theta_D\hspace{-0.25ex})}} }\hspace{-0.25ex},\hspace{-0.25ex} \cdots,\hspace{-0.15ex}e^{\hspace{-0.15ex} -j2\pi d (\hspace{-0.25ex}M_D -1\hspace{-0.25ex}) {{\cos (\theta_D\hspace{-0.25ex})}} } \hspace{-0.25ex}]$}\hspace{-0.75ex} \in\hspace{-0.75ex} \mathbb{C}^{\hspace{-0.25ex}M_D \hspace{-0.25ex}\times\hspace{-0.25ex} 1\hspace{-0.25ex}}, \hspace{2ex}\\ 
	\mathbf{\Phi\hspace{-0.25ex}}_U \hspace{-0.25ex}(\hspace{-0.25ex}\theta_U\hspace{-0.25ex}) \hspace{-0.65ex}&\hspace{-1ex}=\hspace{-1ex}&\hspace{-0.15ex} \mbox{$\squeezespaces{0.1}\hspace{-0.75ex} [\hspace{-0.25ex}1,\hspace{-0.15ex}e^{\hspace{-0.15ex} j2\pi d {{\cos (\theta_U\hspace{-0.25ex})}} }\hspace{-0.25ex},\hspace{-0.25ex} \cdots,\hspace{-0.15ex}e^{\hspace{-0.15ex} j2\pi d (\hspace{-0.25ex}M_U -1\hspace{-0.25ex}) {{\cos (\theta_U\hspace{-0.25ex})}} } \hspace{-0.15ex}]$}\hspace{-0.75ex} \in\hspace{-0.75ex} \mathbb{C}^{\hspace{-0.25ex}M_U \hspace{-0.25ex}\times\hspace{-0.25ex} 1\hspace{-0.25ex}}.
\end{eqnarray}
Let $\hat{\theta}_D$ and $\hat{\theta}_U$ are the beamsteering azimuth angles of the Tx and Rx beams, respectively. Then, the Tx and Rx RF beamformers (beamsteering vectors) can be written as follows:
\begin{eqnarray}
	\mathbf{f\hspace{-0.25ex}}_D \hspace{-0.25ex}(\hspace{-0.25ex}\hat{\theta}_D\hspace{-0.25ex}) \hspace{-0.75ex}& \hspace{-1.5ex}=\hspace{-0.75ex}\frac{1}{\sqrt{M_D}}\hspace{-1.5ex}&\hspace{-0.15ex} \mbox{$\squeezespaces{0.1}\hspace{-0.75ex} [\hspace{-0.25ex}1\hspace{
			-0.2ex},\hspace{-0.25ex}e^{\hspace{-0.25ex}j2\pi d {{\cos (\hat{\theta}_D\hspace{-0.25ex})}} }\hspace{-0.25ex},\hspace{-0.25ex} \cdots,\hspace{-0.25ex}e^{\hspace{-0.25ex}j2\pi d (\hspace{-0.25ex}M_D -1\hspace{-0.25ex}) {{\cos (\hat{\theta}_D\hspace{-0.25ex})}} } \hspace{-0.25ex}]^T$}\hspace{-0.75ex} \in\hspace{-0.75ex} \mathbb{C}^{\hspace{-0.25ex}M_D \hspace{-0.25ex}\times\hspace{-0.25ex} 1\hspace{-0.25ex}}, \hspace{2ex} \label{eq:F_D}\\ 
	\mathbf{f\hspace{-0.25ex}}_U \hspace{-0.25ex}(\hspace{-0.25ex}\hat{\theta}_U\hspace{-0.25ex}) \hspace{-0.65ex}&\hspace{-1.5ex}=\hspace{-1ex}\frac{1}{\sqrt{M_U}}\hspace{-1.3ex}&\hspace{-0.15ex} \mbox{$\squeezespaces{0.1}\hspace{-0.9ex} [\hspace{-0.25ex}1\hspace{-0.25ex},\hspace{-0.25ex}e^{\hspace{-0.85ex} -j2\pi d {{\cos (\hat{\theta}_U\hspace{-0.25ex})}} }\hspace{-0.3ex},\hspace{-0.25ex} \cdots,\hspace{-0.3ex}e^{\hspace{-0.85ex} -j2\pi d (\hspace{-0.25ex}M_U -1\hspace{-0.25ex}) {{\cos (\hat{\theta}_U\hspace{-0.25ex})}} } \hspace{-0.25ex}]^{T}$}\hspace{-1.55ex} \in\hspace{-0.75ex} \mathbb{C}^{\hspace{-0.25ex}M_U \hspace{-0.25ex}\times\hspace{-0.25ex} 1\hspace{-0.25ex}}. \label{eq:F_U}
\end{eqnarray}
\begin{figure}[!t] 
	\centering
	\subfloat[\label{fig:fig3a}]{%
		\includegraphics[height=3.1cm, width=1\columnwidth]{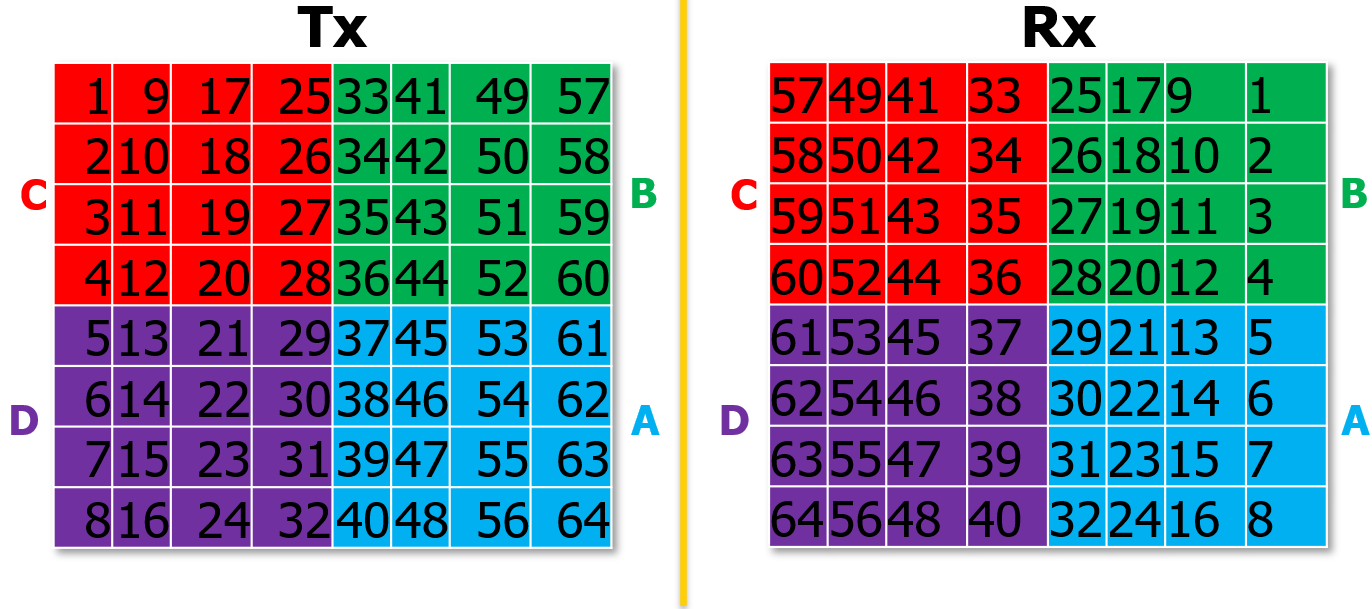} 
	} \hfil 
	\subfloat[\label{fig:fig3b}]{%
		\includegraphics[height=3.1cm, width=1\columnwidth]{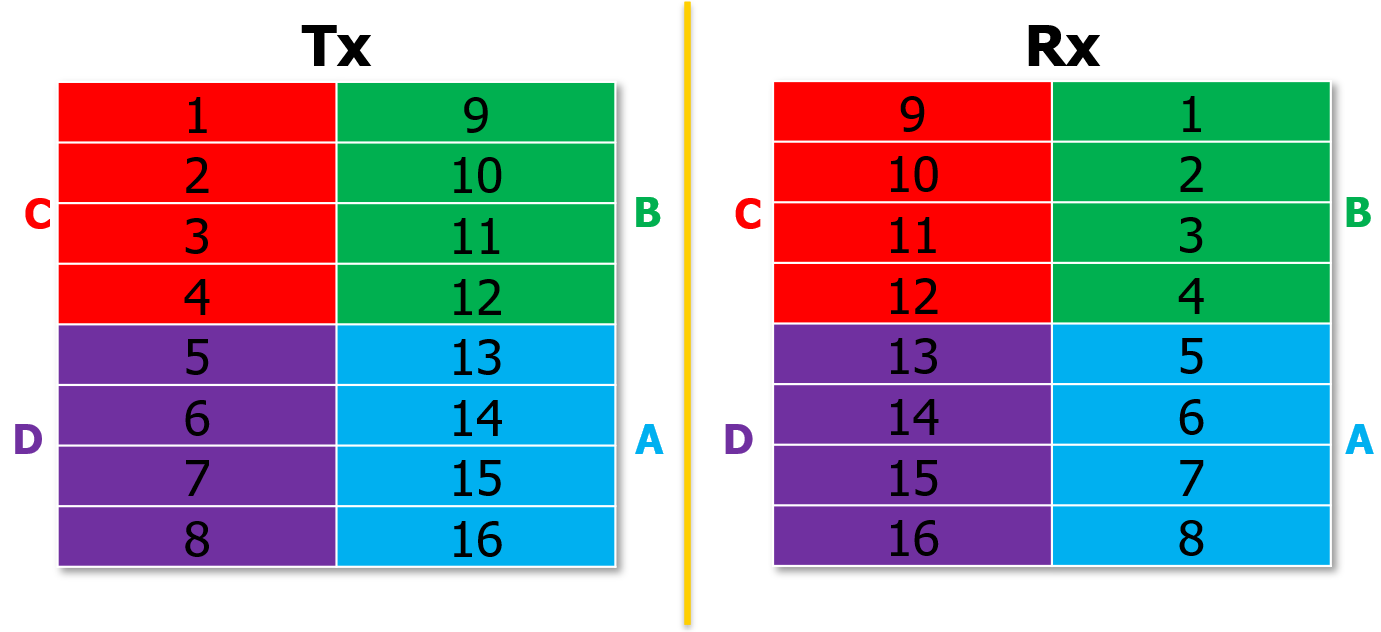} 
	}  \hfil 
	\subfloat[\label{fig:fig3b}]{%
		\includegraphics[height=3.1cm, width=1\columnwidth]{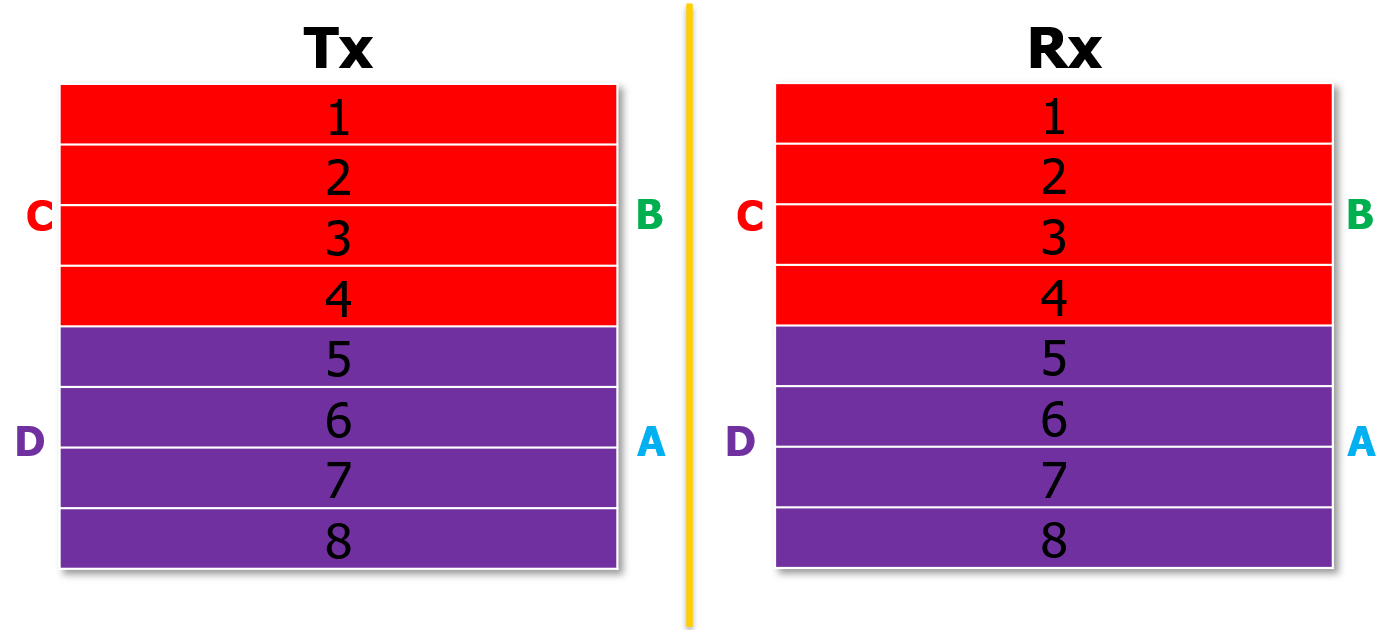} 
	}   
	\caption{Antenna mapping. (a) 64 Tx and 64 Rx antennas index. (b) 1$\times$4 Tx and Rx sub-array mapping. (c) 1$\times$8 Tx and Rx sub-array mapping.}
	\label{fig:fig3} 
	\vspace{-2em}
\end{figure}
\vspace{-2ex}
\subsection{SI Channel Measurement Setup}
The measurement setup was made in an anechoic chamber (i.e., without external surrounding reflections) and consists of 64 Tx and 64 Rx antenna elements\footnote{Due to limited space, the details of SI channel measurement setup will be discussed in the extended version of this paper.}, which are arranged in the form of URA ($8$$\times$$8$ configuration) as shown in Fig. \ref{fig:fig2}. The SI channel is mainly due to \textit{internal} coupling between Tx and Rx antenna elements (i.e., consisting of only line-of-sight (LoS) path components). Then, the SI channel is measured for 1601 sampling points between frequency range from 3 GHz to 4 GHz (i.e., a bandwidth (BW) of 1 GHz) such that the complete SI channel matrix $\mathbf{H}_{SI,ALL}$ has dimensions of 64 $\times$ 64 $\times$ 1601. Particularly, we consider the linear sub-array configurations of 4 and 8 antenna elements for both Tx and Rx as shown in Fig. \ref{fig:fig3}. Hence, the corresponding SI channels for 1$\times$4 and 1$\times$8 sub-array configurations can be represented as $\mathbf{H}_{SI}$ $\in$ $\mathbb{C}^{4 \times 4 \times 1601}$ and $\mathbf{H}_{SI}$ $\in$ $\mathbb{C}^{8 \times 8 \times 1601}$, respectively. As per 3GPP specification, the UL and DL channel BW can vary from 5 MHz to 100 MHz \cite{3GPP_BW}, then the corresponding SI channel for the given BW can be written as: $\mathbf{H}_{SI,B}$ = $\mathbf{H}_{SI}(:,:,n)$ $\in$ $\mathbb{C}^{i \times i \times n}$, where $i = \{4,8\}$, $B$ is the given BW, and $n = 1,2,\dots, N$ is the sample frequency point selected from a total of $N$ frequency points for a given BW. For instance, for a BW of 20 MHz, $n =1,2,\dots,33$ for the frequency range from 3.49 GHz to 3.51 GHz. Similarly, for a BW of 100 MHz, $n =1,2,\dots,161$ for the frequency range from 3.45 GHz to 3.55 GHz. Then, based on the DL and UL RF beamforming stages, we can write the total achieved SI as follows:
\begin{equation}
	\textrm{A}_{\textrm{SI}}\hspace{-0.5ex} =\hspace{-0.5ex} -10\log_{10}\hspace{-0.6ex}\Big(\frac{1}{N}\sum\nolimits_n\hspace{-0.6ex} \big|\hspace{-0.25ex}\mathbf{f}_U^T(\hat{\theta}_U)\mathbf{H}_{SI}(:,:,n)\mathbf{f}_D(\hat{\theta}_D) \hspace{-0.25ex}\big|^2\Big). \label{eq:SIC}
\end{equation}
If we steer the UL and DL beams to the desirable directions (i.e. $\hat{\theta}_U = \theta_U$, $\hat{\theta}_D = \theta_D$),  then the DL and UL directivities are the maximum, which are given as follows:
\begin{equation}
	|\mathbf{\Phi}_D^T(\theta_D)\mathbf{f}_D(\hat{\theta}_D)|^2\hspace{-0.3ex} =\hspace{-0.3ex} M_D, \hspace{1em} |\mathbf{f}_U^T(\hat{\theta}_U)\mathbf{\Phi}_U(\theta_U)|^2 \hspace{-0.3ex}= \hspace{-0.3ex} M_U. 
\end{equation}
For a FD mMIMO system consisting of DL and UL RF beamformers $\mathbf{f}_D$ and $\mathbf{f}_U$, and using sub-array structures at Tx and Rx of BS, for instance, 1$\times$4 or 1$\times$8. Then, the SI can be minimized by the joint optimization of UL and DL perturbation angles $\hat{\theta}_U$, $\hat{\theta}_D$ together with Tx/Rx sub-array selection. Let $i$ and $j$ represents the sub-array indices for Tx and Rx, respectively, then, we can formulate the optimization problem for min-SI hybrid BF under directivity degradation constraints as follows:
\begin{gather}
	\begin{align} 
		&\min_{\left\{ \hat{\theta}_D, \hat{\theta}_U, i, j \right\}} \quad \frac{1}{N}\sum\nolimits_n \big|\mathbf{f}_U^T(\hat{\theta}_U)\mathbf{H}_{SI,i,j}(:,:,n)\mathbf{f}_D(\hat{\theta}_D) \big|^2 \notag \\
		&\hspace{2em}\textrm{s.t.} \hspace{2ex} C_1: \hspace{1ex}  M_D -|\mathbf{\Phi}_D^T(\theta_D)\mathbf{f}_D(\hat{\theta}_D)|^2 \leq \epsilon, \notag \\
		&\hspace{2em} \hspace{4.5ex} C_2: \hspace{1ex}  M_U-|\mathbf{f}_U^T(\hat{\theta}_U)\mathbf{\Phi}_U(\theta_U)|^2 \leq \epsilon, 
	\end{align} 
\end{gather} 
where $C_1$ and $C_2$ refer to the directivity degradation constraints in DL and UL directions, respectively. In other words, we limit the degradation of directivities from the main beam directions $\theta_D$ and $\theta_U$ to a small value $\epsilon$. The optimization problem defined in (9) is non-convex and intractable due to the non-linearity constraints.
\vspace{-1ex}
\section{Tx and Rx Sub-Array Mapping and Proposed Joint Min-SI BF and SAS}
\vspace{-1ex}
In this section, our objectives are to suppress strong SI solely based on the design of min-SI RF-BF stages $\mathbf{f}_U$ and $\mathbf{f}_D$ jointly with SAS to provide an additional DoF in FD mMIMO systems, which can avoid the use of costly analog cancellation circuits. In Fig. \ref{fig:fig3}(a), the antenna mapping is shown for both Tx and Rx of BS, which consists of 64 elements at BS and separated by an antenna isolation block. At first, we discuss the sub-array mapping for our given Tx/Rx setup.
\vspace{-1ex}
\subsection{Sub-Array Mapping}
\vspace{-1ex}
We consider the following two different sub-array configurations for Tx and Rx: 1) 1$\times$4 sub-array; and 2) 1$\times$8 sub-array. Given 64 Tx or Rx antenna elements, we can have 16 possible Tx and Rx sub-arrays of 1$\times$4 elements, which are arranged in the form of ULA. Fig. \ref{fig:fig3}(b) depicts the mapping of 16 different 1$\times$4 sub-arrays for both Tx and Rx. For instance, sub-array 1 for Tx and Rx constitutes antenna elements with index values 1,9,17,25. It can be seen that using 1$\times$4 sub-arrays at Tx and Rx can give rise to $16\times16 = 256$ possible combinations for the Tx and Rx sub-array selection, which can be computationally expensive. Similarly, Fig. \ref{fig:fig3}(c) shows the mapping for different 1$\times$8 sub-arrays for both Tx and Rx. For instance, sub-array 1 for Tx and Rx now constitutes antennas with indices 1,9,17,25,33,41,49,57. The selection of 1$\times$8 Tx and 1$\times$8 Rx sub-array gives rise to $8\times8 = 64$ possible combinations for SAS.
\vspace{-1ex}
\subsection{Min-SI BF with SAS}
\vspace{-1ex}
We propose a particle swarm optimization (PSO)-based min-SI BF with SAS scheme to find the optimal DL and UL beam directions $\hat{\theta}_D, \hat{\theta}_U$ together with Tx sub-array index $i$ and Rx sub-array index $j$ to minimize SI while satisfying the corresponding directivity degradation constraints $C_1$ and $C_2$. The algorithm starts with a swarm of $N_p$ particles, each with its own position, velocity, and fitness value, which are randomly placed in optimization search space of perturbation coefficients. During a total of $T$ iterations, the particle $p$ communicates with each other, and move for the exploration of the optimization space to find the optimal solution. Here, we define the perturbation vector $\mathbf{X}_p^{(t)}$ as follows:
\begin{equation}
	\mathbf{X}_p^{(t)} = [\hat{\theta}_D^p, \hat{\theta}_U^p, i^p, j^p], \label{eq:PSO}
\end{equation}
where $p = 1,\dots, N_p$ and $t = 0,1, \dots, T$. For each $p^{th}$ particle, by substituting (\ref{eq:PSO}) in (\ref{eq:F_D}) and (\ref{eq:F_U}), the DL and UL RF beamformers $\mathbf{f}_D(\mathbf{X}_p^{(t)})$ and $\mathbf{f}_U(\mathbf{X}_p^{(t)})$ can be obtained as function of perturbation angles $\hat{\theta}_D^p$ and $\hat{\theta}_U^p$, respectively.
\begin{algorithm}[t!]\label{algo:3}
	\nonl \textbf{Input}: \hspace{-0.5ex} $N_p, T$, $\mathbf{H}_{SI}$, $(\theta_D, \psi_D)$, $(\theta_U, \psi_U)$. \\
	\nonl \textbf{Output}: \hspace{-0.5ex}$ \hat{\theta}_D, \hat{\theta}_U, i, j$. \\
	\SetAlgoLined 
	\For{$t = 0:T$}{
		\For{$p = 1:N_p$}{
			\uIf{t = 0}{
				Initialize the velocity as $\mathbf{v}_{p}^{(0)} = \bf{0}$. \\
				Initialize $\mathbf{X}_p^{(t)}$ uniformly distributed in $[\mathbf{X}_{\text{Low}}, \mathbf{X}_{\text{Upp}}]$. \\}
			\Else {
				Update the velocity $\mathbf{v}_{p}^{(t)}$ via (\ref{eq:velocity}).\\
				Update the perturbation $\mathbf{X}_{p}^{(t)}$ via (\ref{eq:position}).\\
			}
			Find the personal best $\mathbf{X}_{p,\mathrm{best},n}^{(t)}$ via (\ref{eq:personal_best_PSOLPA}).
		}
		Find the global best $\mathbf{X}_{\text{best}}^{(t)}$ as in (\ref{eq:global_best_PSOLPA}).
	}
	\caption{Min-SI BF with SAS Algorithm} 
\end{algorithm}
\setlength{\textfloatsep}{0pt} \hspace{-1ex}
By using (\ref{eq:SIC}), we can write the achieved SI suppression as follows:
\begin{equation}
	\text{A}_{\textrm{SI}}\hspace{-0.2ex}(\mathbf{X}_p^{(t)}\hspace{-0.5ex}) \hspace{-0.75ex}=\hspace{-0.75ex} -10\hspace{-0.25ex}\log_{10\hspace{-0.25ex}}\hspace{-0.6ex}\Big(\hspace{-0.5ex}\frac{1}{N}\hspace{-0.6ex}\sum_n\hspace{-0.4ex} \big|\hspace{-0.3ex}\mathbf{f}_U^T(\mathbf{X}_p^{(t)}\hspace{-0.35ex})\mathbf{H}_{SI\hspace{-0.35ex}}(\hspace{-0.35ex}\mathbf{X}_p^{(t)\hspace{-0.15ex}}\hspace{-0.35ex})\mathbf{f}_D(\mathbf{X}_p^{(t)}\hspace{-0.35ex})\hspace{-0.3ex}\big|^2\hspace{-0.5ex}\Big). \label{eq:SIC_PSO}
\end{equation}
At the $t^{th}$ iteration, the personal best for the $p^{th}$ particle and the current global best among all particles are respectively found as follows:
\begin{equation}
	\mathbf{X}_{\mathrm{best},p}^{(t)}= \argmin_{\mathbf{X}_{p}^{(t^*)}, \forall t^* = 0,1,\cdots, t} \text{A}_{\textrm{SI}}(\mathbf{X}_p^{(t^*)}), \label{eq:personal_best_PSOLPA}  
	\vspace{-1ex}
\end{equation}
\begin{equation}
	\mathbf{X}_{\mathrm{best}}^{(t)}= \argmin_{\mathbf{X}_{\text{best},p}^{(t)}, \forall p = 0,1,\cdots, N_p} \text{A}_{\textrm{SI}}(\mathbf{X}_{\text{best},p}^{(t)}). \label{eq:global_best_PSOLPA} 
	\vspace{-1ex}
\end{equation}
The convergence of the proposed PSO-based min-SI BF with SAS for enhanced SI suppression depends on the velocity vector $\mathbf{v}_p$ for both personal best $\mathbf{X}_{\mathrm{best},p}$ and global best $\mathbf{X}_{\mathrm{best}}$ solutions, which is defined as follows:
\begin{equation}
	\mathbf{v}_p^{(t+1)}\hspace{-0.65ex}=\hspace{-0.25ex}\mathbf{\Omega}_1(\hspace{-0.25ex}\mathbf{X}_{\mathrm{best}}^{(t)} \hspace{-0.5ex}-\hspace{-0.5ex} \mathbf{X}_{p}^{(t)})\hspace{-0.35ex}+\hspace{-0.25ex}\mathbf{\Omega}_2(\mathbf{X}_{\mathrm{best},p}^{(t)}\hspace{-0.5ex} -\hspace{-0.5ex} \mathbf{X}_{p}^{(t)})+\mathbf{\Omega}_3^{(t)} v_p^{(t)}, \label{eq:velocity} \vspace{-1ex}
\end{equation}
where $\mathbf{v}_p^{(t)}$ is the velocity of the $p^{th}$ particle at the $t^{th}$ iteration, $\mathbf{\Omega}_1, \mathbf{\Omega}_2$ are the random diagonal matrices with the uniformly distributed entries over $[0,2]$ and represent the social relations among the particles, and the tendency of a given particle for moving towards its personal best, respectively. Here, $\mathbf{\Omega}_3=\left(\frac{T-1}{T}\right) \mathbf{I}_{\left(2 N_D + 2 N_U\right)}$ is the diagonal inertia weight matrix, which finds the balance between exploration and exploitation for optimal solution in search space. By using (\ref{eq:velocity}), the position of each particle during $t^{th}$ iteration is updated as:
\begin{equation}
	\mathbf{X}_p^{(t+1)}=\operatorname{clip}\left(\mathbf{X}_p^{(t)} + \mathbf{v}_p^{(t+1)}, \mathbf{X}_{\text{Low}}, \mathbf{X}_{\text{Upp}}\right), \label{eq:position} \vspace{-1ex}
\end{equation}
where $\mathbf{X}_{\text{Low}} \in \mathbb{R}^{\left(2 N_D+2 N_U\right)}$ and $\mathbf{X}_{\text{Upp}} \in \mathbb{R}^{\left(2 N_D+2 N_U\right)}$ are the lower-bound and upper-bound vectors for the perturbation coefficients, respectively, and are constructed according to the earlier defined boundaries of each perturbation coefficient given in $C_1$ and $C_2$. Here, we define $\operatorname{clip}(x, a, b)=$ $\min (\max (x, a), b)$ as the clipping function to avoid exceeding the bounds. Furthermore, different from the sub-optimal approach, we here consider each perturbation coefficient as a continuous variable inside its boundary. The proposed perturbation-based SI minimization with SAS scheme using PSO is summarized in Algorithm 1. 
\begin{figure}[!t] 
	\centering
	\subfloat[\label{fig:fig51a}]{%
		\includegraphics[scale = 0.2]{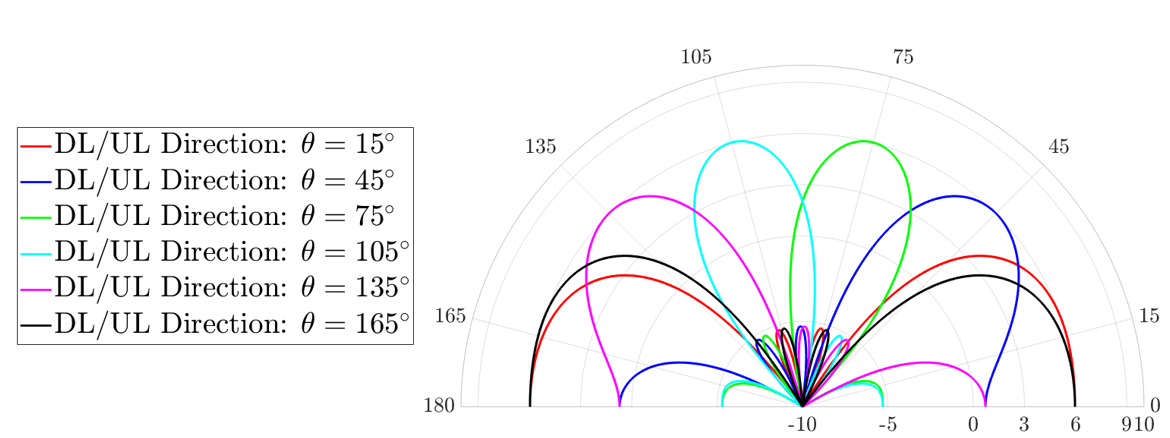} 
	} \hfil 
	\subfloat[\label{fig:fig52b}]{%
		\includegraphics[scale = 0.2]{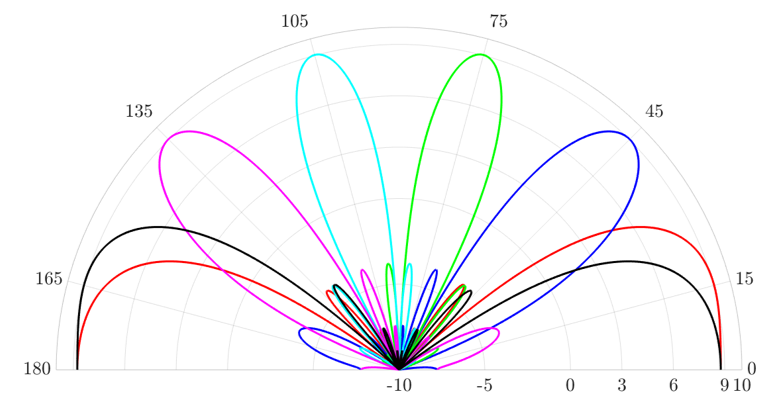} 
	}   
	\caption{Beampatterns via DBF. (a) 1$\times$4 sub-array. (b) 1$\times$8 sub-array.}
	\label{fig:fig55} 
\end{figure}
\begin{figure}[!t] 
	\centering
	\subfloat[\label{fig:fig41a}]{%
		\includegraphics[scale = 0.25]{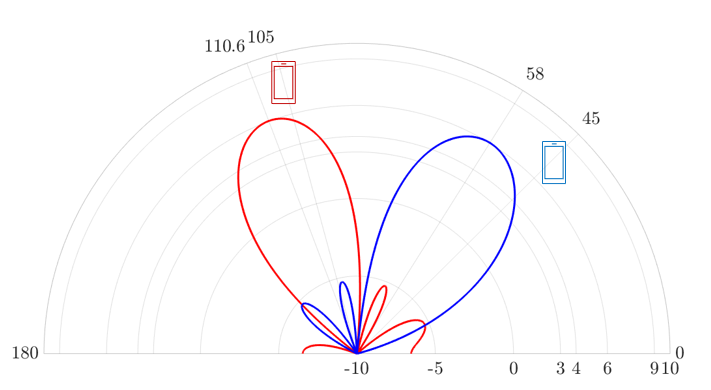} 
	} \hfil 
	\subfloat[\label{fig:fig42b}]{%
		\includegraphics[scale = 0.24]{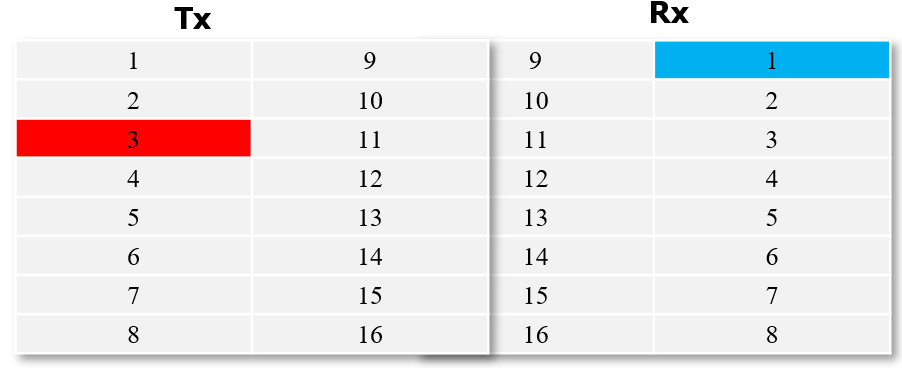} 
	}   
	\caption{Proposed min-SI BF with SAS. (a) UL and DL beam perturbations. (b) Tx and Rx sub-array indices for SAS.}
	\label{fig:fig44} 
	\vspace{2ex}
\end{figure}
\section{Illustrative Results}
In this section, we present the Monte Carlo simulation results to illustrate the performance of the proposed SI suppression technique in FD mMIMO systems. Particularly, we investigate the amount of achieved SI suppression by the design of min-SI RF-BF stages together with SAS. We consider $N_D$ $=$ $N_U$ $=$ 1 RF chain to serve a single UL and DL UE with 1$\times$4 and 1$\times$8 sub-array configurations for the results presented hereafter. For PSO, we use $N_p=20, \Omega_1=\Omega_2=2$ and $\Omega_3=1.1$. In Fig. \ref{fig:fig55}, we plot the beampatterns using both 1$\times$4 and 1$\times$8 sub-arrays for six different angular locations of UL/DL UE (i.e., $\{ \theta_D, \theta_U\} \in \{15^{\circ}:30^{\circ}:180^{\circ}\}$). In particular, we refer the case when the beams generated by the UL and DL RF beamformers are steered at exact UE locations (i.e., both $\mathbf{f}_D(\theta_D)$ and $\mathbf{f}_U(\theta_U)$ steer the beams at $\theta_D$ and $\theta_U$, respectively) as directivity-based beamforming (DBF). It can be seen that 1$\times$8 sub-array can generate narrower beams when compared to 1$\times$4 sub-array, and can serve more number of users when compared to 1$\times$4 sub-array. However, due to the orthogonality, there is still a limitation on the number of the orthogonal UL/DL beams that can be generated with 1$\times$8 sub-array. As a result, using DBF restricts the maximum number of UL and DL users that can be served simultaneously in FD mMIMO systems. In the following, we present SI suppression results for 1$\times$4 and 1$\times$8 sub-array configurations using the proposed min-SI BF with SAS scheme.
\subsection{SI Suppression Using 1$\times$4 Sub-Array}
Fig. \ref{fig:fig44} presents the results using min-SI BF with SAS for 1$\times$4 sub-array for both Tx and Rx. We consider DL and UL UE located at angular locations $\theta_D = 105^{\circ}$ and $\theta_U = 45^{\circ}$, respectively. It must be noted that compared to DBF RF beamformers $\mathbf{f}_D(\theta_D)$ and $\mathbf{f}_U(\theta_U)$, which direct the beams in the desired UE directions $\theta_D$, $\theta_U$, the min-SI RF beamformers with SAS introduce beam perturbations at $\hat{\theta}_D$ and $\hat{\theta}_U$  (i.e., $\mathbf{f}_D(\hat{\theta}_D)$ and $\mathbf{f}_U(\hat{\theta}_U)$). The proposed scheme then finds the optimal perturbations as $110.6^{\circ}$ and $58^{\circ}$ for DL and UL beams, respectively. Moreover, as shown in Fig. \ref{fig:fig44}(b), the optimal Tx and Rx sub-array indices are found to be 3 and 1, respectively which can achieve SI suppression of around 78.5 dB at the expense of directivity degradation of $\epsilon=2$ dB.\par 
\begin{figure}[!t] 
	\centering
	\subfloat[\label{fig:fig4a}]{%
		\includegraphics[scale = 0.25]{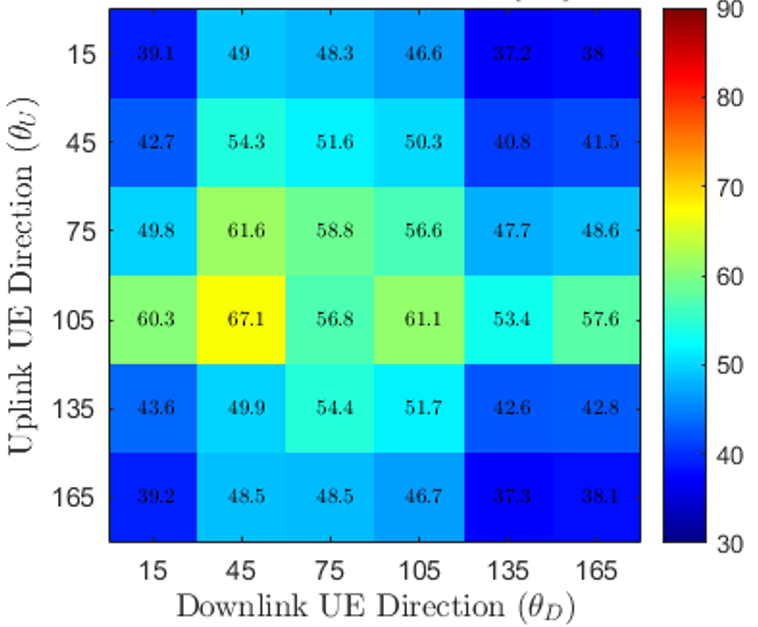} 
	} \hfil 
	\subfloat[\label{fig:fig4b}]{%
		\includegraphics[scale = 0.24]{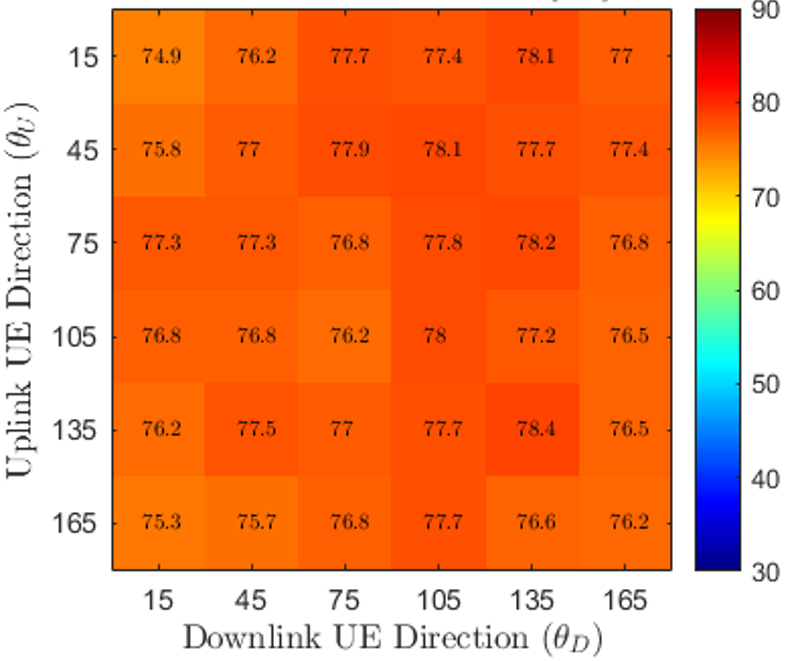} 
	}   
	\caption{SI suppression using $1\times4$ sub-array at 20 MHz. (a) DBF. (b) min-SI BF with SAS.}
	\label{fig:fig4} 
	\vspace{1ex}
\end{figure}
\begin{figure}[!t] 
	\centering
	\subfloat[\label{fig:fig5a}]{%
		\includegraphics[scale = 0.25]{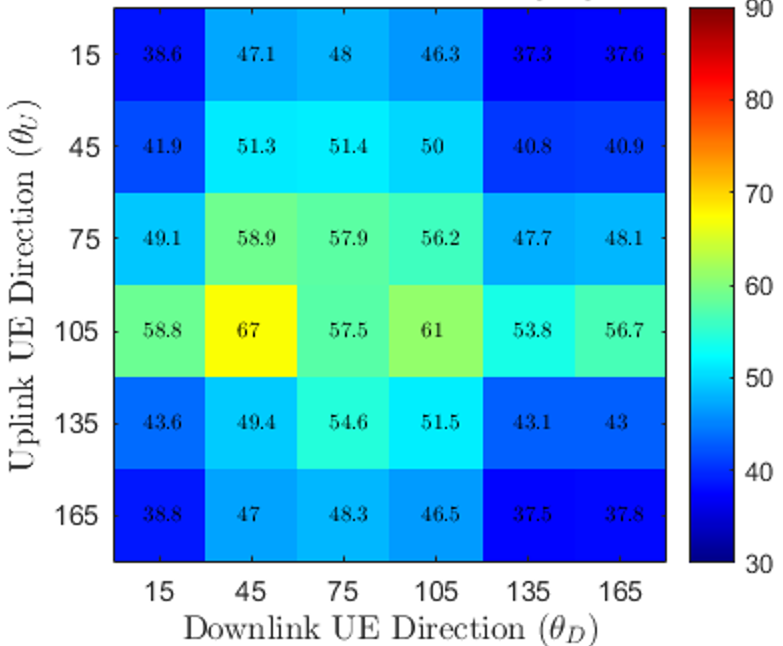} 
	} \hfil 
	\subfloat[\label{fig:fig5b}]{%
		\includegraphics[scale = 0.25]{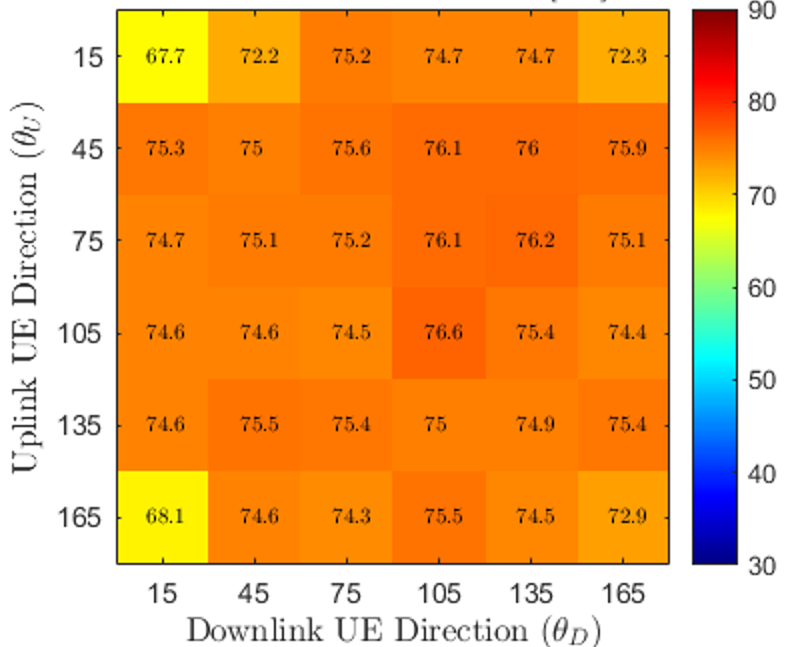} 
	}   
	\caption{SI suppression using 1$\times$4 sub-array at 100 MHz. (a) DBF. (b) min-SI BF with SAS.}
	\label{fig:fig5} 
		\vspace{1ex}
\end{figure}
In Fig. \ref{fig:fig4}, we compare the achieved SI using BW = 20 MHz for the following two schemes: 1) proposed min-SI BF with SAS; and 2) DBF. We consider 6 different angular locations for UL and DL UE (i.e., $\{\theta_D, \theta_U\} \in [15^{\circ}:30^{\circ}:180^{\circ}]$). It can be seen that the design of RF beamformers $\mathbf{f}_D(\theta_D)$ and $\mathbf{f}_U(\theta_U)$ using DBF can achieve SI suppression ranging from 37.2 dB to 67.1 dB for different UL/DL UE angle pairs. On the other hand, the proposed min-SI BF scheme with SAS can achieve SI suppression ranging from 74.9 dB to 78.4 dB. This shows that the design of min-SI RF beamformers $\mathbf{f}_D(\hat{\theta}_D)$, $\mathbf{f}_U(\hat{\theta}_U)$ with SAS can provide an additional SI gain of 33 dB on average when compared to DBF, and can improve SI suppression by a maximum of 40.1 dB (e.g., for $\theta_D= 135^{\circ}, \theta_U= 15^{\circ}$, SI suppression improves from 37.2 dB to 78.1 dB.). Similarly, Fig. \ref{fig:fig5} shows the enhanced SI suppression for the proposed min-SI BF scheme with SAS using BW = 100 MHz. DBF can provide SI suppression ranging from 37.3 dB to 67 dB, whereas, the proposed min-SI BF with SAS can achieve SI suppression ranging from 67.7 dB to 76.6 dB. Thus, the proposed min-SI BF scheme can provide an SI suppression gain of around 30.3 dB on average with a maximum SI suppression gain of 37.4 dB.  
\subsection{SI Suppression Using 1$\times$8 Sub-Array}
\begin{figure}[!t] 
	\centering
	\subfloat[\label{fig:fig6a}]{%
		\includegraphics[scale = 0.24]{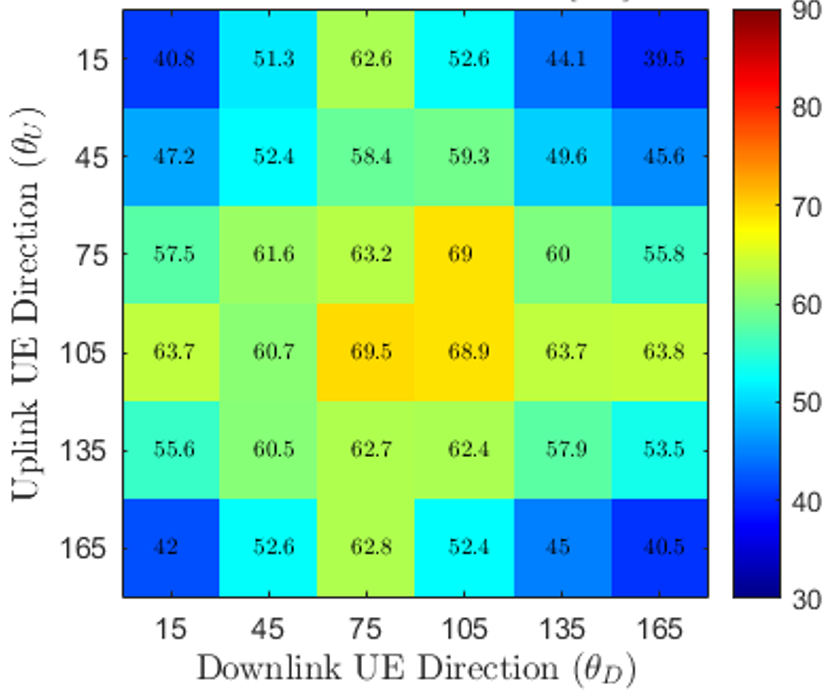} 
	} \hfil 
	\subfloat[\label{fig:fig6b}]{%
		\includegraphics[scale = 0.24]{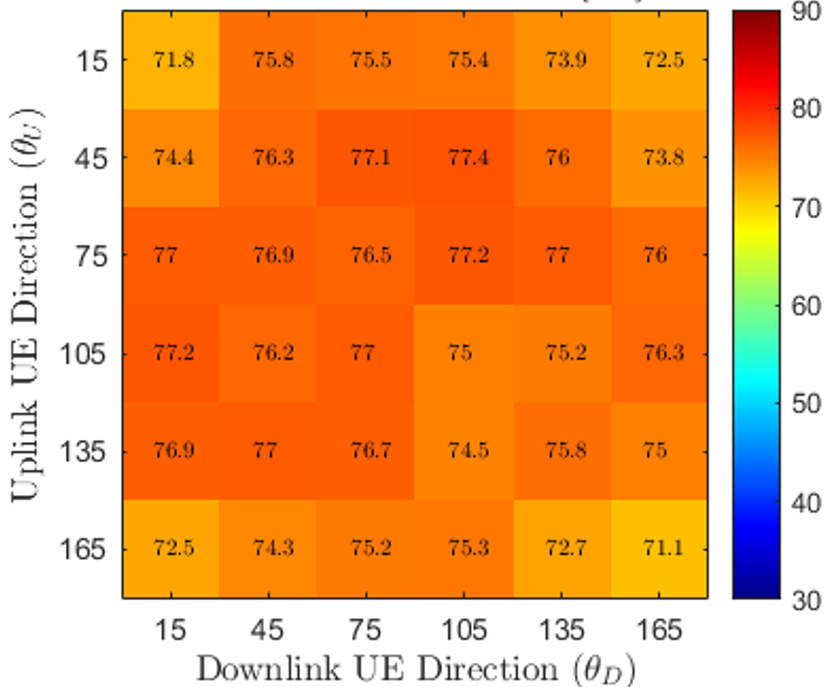} 
	}   
	\caption{SI suppression using $1\times8$ sub-array at 20 MHz. (a) DBF. (b) min-SI BF with SAS.}
	\label{fig:fig6} 
\end{figure}

\begin{figure}[!t] 
	\centering
	\subfloat[\label{fig:fig7a}]{%
		\includegraphics[scale = 0.25]{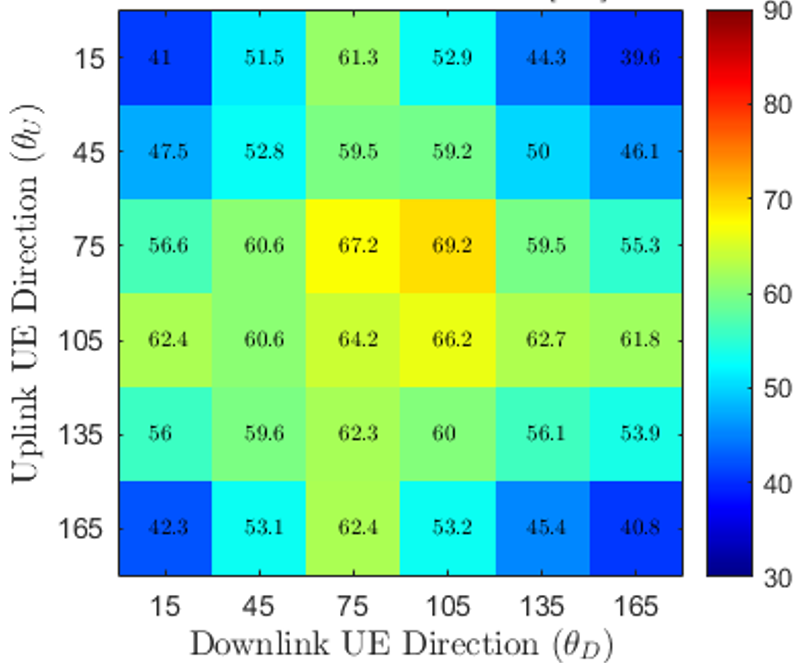} 
	} \hfil 
	\subfloat[\label{fig:fig7b}]{%
		\includegraphics[scale = 0.25]{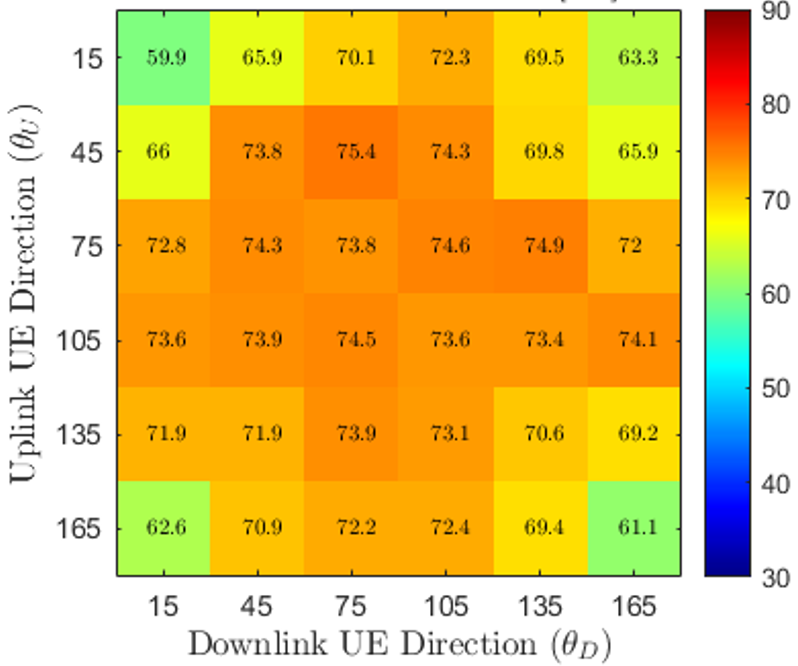} 
	}   
	\caption{SI suppression using 1$\times$8 sub-array at 100 MHz. (a) DBF. (b) min-SI BF with SAS.}
	\label{fig:fig7} 
		\vspace{1ex}
\end{figure}

In this section, we present the results by using 1$\times$8 sub-array for both Tx and Rx for a FD mMIMO system. Fig. \ref{fig:fig6} depicts the SI suppression results using the proposed min-SI BF approach with SAS for 6 different UL and DL angular locations using BW $=$ 20 MHz. The use of a larger array structure can further suppress SI by generating narrower beams. Therefore, compared to SI suppression ranging between 39.5 dB and 69.5 dB for DBF, the proposed min-SI BF scheme can achieve SI suppression ranging from 71.1 dB to 77.4 dB. On average, the proposed scheme can provide an SI suppression gain of around 24.8 dB with a maximum suppression gain of 33 dB at $\theta_D= 165^{\circ}, \theta_U= 15^{\circ}$. Fig. \ref{fig:fig7} depicts the achieved SI suppression using BW = 100 MHz. By designing $\mathbf{f}_D(\theta_D)$ and $\mathbf{f}_U(\theta_U)$ using DBF can achieve SI suppression between 39.6 and 69.2 dB, whereas, the proposed min-SI BF scheme can provide SI suppression ranging from 59.9 dB to 75.4 dB. Thus, the use of min-SI BF with SAS can provide an additional SI suppression of around 17.9 dB and a maximum SI suppression gain of 25.2 dB. Compared to SI suppression results with 20 MHz, a slightly lower SI suppression is achieved with BW of 100 MHz due to the use of larger number of frequency sampling points (as given in (\ref{eq:SIC})).
\section{Conclusions}
In this paper, we have considered a novel FD mMIMO systems using HBF architecture for simultaneous UL and DL transmission over the same frequency band. In particular, we have addressed the optimization problem of suppressing the strong SI solely on the design of UL and DL RF beamforming stages jointly with Tx and Rx SAS. Based on the measured SI channel, we have proposed a novel min-SI BF scheme jointly with SAS for both Tx and Rx sub-arrays. To solve this challenging non-convex problem, we have proposed a swarm intelligence-based algorithmic solution to find the optimal perturbations as well as the Tx and Rx sub-arrays while satisfying the directivity degradation constraints for the UL and DL beams. The results show that min-SI BF scheme together with SAS can achieve high SI suppression when compared to DBF for both 1$\times$4 and 1$\times$8 sub-array configurations, and can achieve SI suppression as high as 78 dB for FD mMIMO systems.

\balance
\bibliographystyle{IEEEtran}
\bibliography{references}

\end{document}